\documentclass[5p]{elsarticle}

\usepackage{hyperref}
\usepackage{siunitx}
\usepackage{cleveref}
\usepackage{xcolor}
\usepackage{nicefrac}

\journal{Astronomy and Computing}

\begin{document}

\begin{frontmatter}

\title{A Machine Learning Approach to the Detection of Ghosting and Scattered Light Artifacts in Dark Energy Survey Images}


\author[addruchicago,addrkicp]{Chihway~Chang}
\author[addrfnal,addrkicp,addruchicago]{Alex~Drlica-Wagner}
\author[addrfnal]{Stephen~M.~Kent}
\author[addrfnal,addrkicp]{Brian~Nord}
\author[addrimsa]{Donah~Michelle~Wang}
\author[addrfnal]{Michael~H.~L.~S.~Wang\corref{mycorrespondingauthor}}
\cortext[mycorrespondingauthor]{Corresponding author}
\ead{mwang@fnal.gov}

\address[addruchicago]{University of Chicago, Chicago, IL 60637, USA}
\address[addrkicp]{Kavli Institute for Cosmological Physics, University of Chicago, Chicago, IL 60637, USA}
\address[addrfnal]{Fermi National Accelerator Laboratory, Batavia, IL 60510, USA}
\address[addrimsa]{Illinois Mathematics and Science Academy, Aurora, IL 60506, USA}

\begin{abstract}
Astronomical images are often plagued by unwanted artifacts that arise from a
number of sources including imperfect optics, faulty image sensors, cosmic ray
hits, and even airplanes and artificial satellites.  Spurious reflections (known
as ``ghosts'') and the scattering of light off the surfaces of a camera and/or
telescope  are particularly difficult to avoid.  Detecting ghosts and scattered
light efficiently in large cosmological surveys that will acquire petabytes of
data can be a daunting task.  In this paper, we use data from the Dark Energy
Survey to develop, train, and validate a machine learning model to detect ghosts
and scattered light using convolutional neural networks.  The model architecture
and training procedure is discussed in detail, and the performance on the
training and validation set is presented. Testing is performed on data and
results are compared with those from a ray-tracing algorithm. As a proof of
principle, we have shown that our method is promising for the Rubin Observatory
and beyond.
\end{abstract}


\end{frontmatter}


\section{Introduction}

\begin{figure*}[htbp]
\centering
  \begin{tabular}[t]{ccc}
\includegraphics[width=0.31\textwidth]{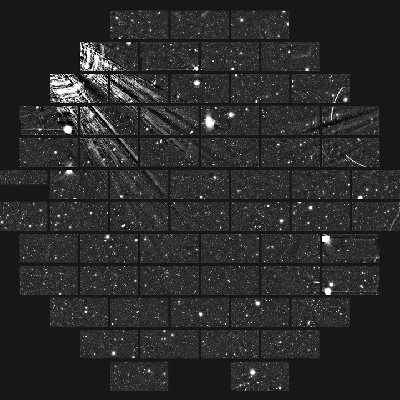} &
\includegraphics[width=0.31\textwidth]{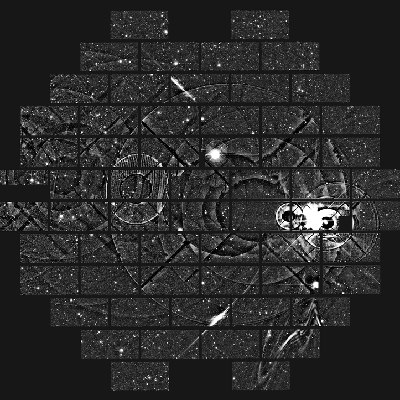} &
\includegraphics[width=0.31\textwidth]{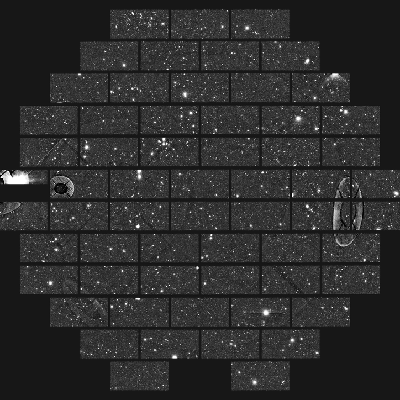} \\
(a) & (b) & (c)\\
  \end{tabular}
\caption{Example full focal plane DECam images that exhibit ghosts and scattered light artifacts.}
\label{fig:ghosts}
\end{figure*}

When the Dark Energy Survey (DES)~\cite{ref:des2005,ref:des2016} completed its
mission in January 2019, it had mapped $\sim$5000 square degrees of the 
southern sky using the 570 megapixel Dark Energy Camera (DECam)
\cite{ref:flaugher2015} mounted on the Blanco 4-m telescope at the Cerro Tololo
Inter-American Observatory in the Chilean Andes. Over the course of 758 nights
of data taking spread across 6 years, DES generated a massive $\sim$2 petabytes
of data.  Due to the nature of the DECam optical systems, the DES data are
subject to imaging artifacts caused by spurious reflections (commonly referred
to as ``ghosts'') and scattered light \cite{ref:steveslides} (see 
Figure~\ref{fig:ghosts}).  While all astronomical objects observed by DECam
produce ghosts and scattered light at some level, this study specifically
focuses on identifying artifacts from bright stars that are prominent enough to
have a negative impact on object detection, background estimation, and
photometric measurements.  In particular, ghosts/scattered light present a major
source of contamination for studies of low-surface-brightness galaxies and
present a major challenge for precision photometry of faint objects
\cite{ref:tanoglidis2020}. Thus, much effort has been devoted to the mitigation
of such effects.  For example, after the DES science verification data set was
collected, light baffles were installed around all the filters to block a
scattered-light path.  After the first year of DES, the cylindrical interior
surfaces near the optical aperture of the filter changer and shutter were
painted with a black, anti-reflective paint. This paint reduced the number of
possible scattered-light paths and improved the quality of subsequent data sets
\cite{ref:flaugher2015,ref:steveslides}. In this article, we seek to identify
residual ghosts and scattered light artifacts in the DES data. We use the term
``ghosts/scattered light'' to broadly refer to all artifacts that result from
spurious reflections and scattered light  without distinguishing between the
various sources of these artifacts. 

Due to the large volume of DES data, the identification of ghosts and scattered
light by eye is impractical.   DES has automated the detection of these
artifacts through the development of a ray-tracing algorithm that combines a
model of the camera optics, the telescope pointing, and the known locations and
brightness of stars to predict the presence and location of ghosts/scattered
light in an exposure (Section \ref{sec:conventional}).  While this
algorithm correctly identifies and localizes a significant number of
ghosts/scattered light artifacts, it is limited by the accuracy of the
optical model, the telescope  telemetry, and external catalogs of bright stars.
Because the ray-tracing algorithm does not use the DES imaging data directly, it
can miss a substantial number of ghosts/scattered light artifacts. There is
clearly a need for more effective methods to address this problem, especially in
light of future cosmic surveys like the Rubin Observatory Legacy Survey of Space
and Time (LSST), which will have a field of view three times as large as DECam
and will acquire $\sim$20 terabytes of data per night ($\sim$60 petabytes over
ten years) \citep{ref:ivezic2008}.

This paper explores the use of modern machine learning (ML) methods as a
potential solution to the problem of efficiently detecting ghosts/scattered
light in large optical imaging surveys.  Though ML methods have been in use for
over half a century~\cite{ref:alsamuel}, we are referring specifically to the
advances in computer vision made in the past two decades.  These advances were
made possible by the confluence of several key factors that included (1) a
deeper understanding of the internal workings of the visual
cortex~\cite{ref:cortex2}, (2) the introduction of convolutional neural networks
(CNNs) inspired by the visual cortex~\cite{ref:lenet5}, (3) the development of
practical techniques to train such networks~\cite{ref:hintontrain}, and (4) the
availability of vastly increased computational power from devices like graphics
processing units (GPUs).

Attempts have been made to apply such ML techniques to the identification of
telescope artifacts.  In an unpublished report, a CNN was found to
significantly outperform a classical ML algorithm (i.e., a support vector
machine) when both were applied to DES images to identify artifacts
belonging to 28 different classes~\cite{ref:morningstar}.  However, in this
study the CNN showed evidence of overfitting, which the authors suggested could
be mitigated with additional training data. Instead of dealing with multiple
classes of artifacts at once, another effort relied on a CNN-based architecture
to identify artifacts caused by cosmic rays in Hubble Space Telescope
images~\cite{ref:deepcr}.  These authors showed that a CNN-based approach
could provide a significant improvement over the current state-of-the-art
method. In our work, we focus specifically on ghosts/scattered light to
demonstrate a proof-of-principle for the viability of modern ML techniques for
this purpose in large cosmological surveys.

\begin{figure*}[htbp]
\centering
\includegraphics[width=0.90\textwidth]{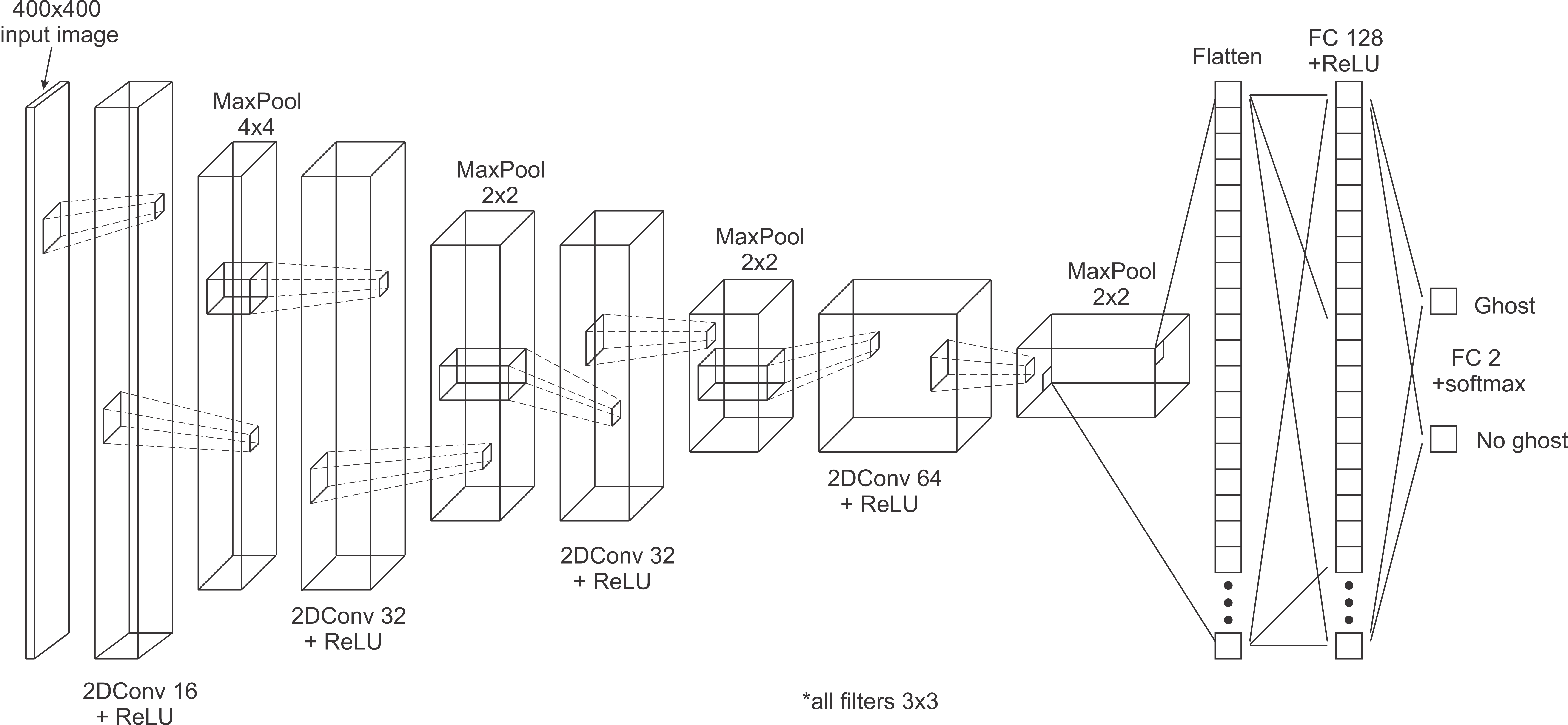}
\caption{Architecture of neural network with four convolutional+maxpool layers followed by two fully connected layers.}
\label{fig:cnn}
\end{figure*}

\section{Conventional Approach}
\label{sec:conventional}
The conventional approach to ghosts/scattered light artifact identification
in DES uses optical ray tracing. A standard optical design program is used to
perform sequential ray tracing to model the performance of the telescope and
optical corrector.   Scattered light comes from grazing incidence scatters off
of surfaces such as the camera filter changer and shutter mechanism
\citep{ref:steveslides}. Ghosts are typically produced by reflections between
two glass surfaces within the corrector, and for each possible combination of
surfaces, ghosts were modeled by introducing two extra mirrored surfaces at the
appropriate positions into the optical design.   The model is quite accurate at
predicting the locations of ghosts, but it has difficulty predicting their
intensities, since those depend on details of reflectivities from antireflection
coatings and filters, which in turn depend on the incidence angle and wavelength
of each ray.   The reflectivities were calibrated empirically from ${\sim} 100$
DES images that contained bright stars of known intensity.   In making
predictions for a validation image, the locations of all known stars were
determined in advance, intensities for all potential ghosts were estimated, and,
if the intensity for a particular ghost exceeded a preset threshold, the area
covered by the ghost was estimated by tracing about 2000 rays sampling the
entrance pupil of the telescope, and all CCDs illuminated by those rays were
flagged as being affected.

While the ray tracing algorithm correctly identifies and localizes a
significant number of ghosts/scattered light artifacts, it is limited by the
accuracy of the optical model and telescope pointing telemetry.  The ray tracing
algorithm also depends on predetermined fluxes of bright stars to predict the
intensity of ghosts/scattered light artifacts. These fluxes are taken from
external catalogs, where they are reported in bands that differ from those
observed by DES. Furthermore, the fluxes of these stars are assumed to be
constant in time, while bright stars are often variable. Because of these
factors, the ray tracing algorithm can miss a substantial number of
ghosts/scattered light artifacts. For this reason, every image that was flagged
by the ray-tracing program was visually inspected, and in some cases, the list
of flagged CCDs was adjusted by hand.

\begin{figure*}[htbp]
\centering
\includegraphics[width=0.9\textwidth]{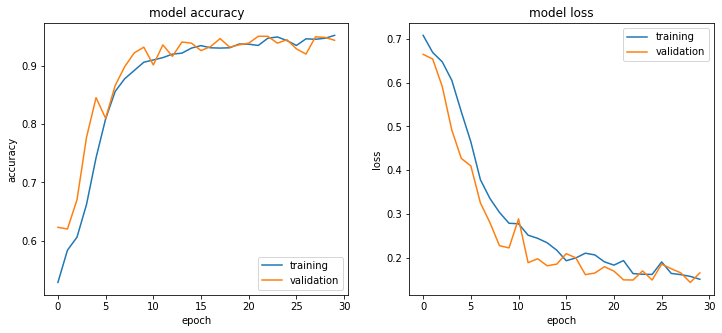}
\caption{Evolution of the accuracy (left) and loss (right) as a function of epoch as evaluated on the training and validation samples.}
\label{fig:accloss}
\end{figure*}


\section{Machine Learning Approach}

Construction, training, and testing of the CNN-based ML model used in this paper were all done using the Tensorflow and Keras machine learning frameworks~\cite{ref:tensorflow2015-whitepaper,ref:chollet2015keras}.

\subsection{Model Architecture}

The choice of network architecture used in this work was guided by our
ultimate goal of investigating whether ML techniques were feasible for detecting
ghosts/scattered light artifacts, and if so, how they would compare with the
conventional technique based on ray tracing.  Since the main objective was a
proof-of-concept demonstration, we opted for a relatively simple CNN
architecture that: (1) was straightforward to implement in a common ML
framework, (2) did not require significant computing resources to train, and (3)
had good performance on standard image classification data sets that would carry
over to artifact detection in DES exposures. The CNN architecture we settled on
was very similar to AlexNet \cite{ref:alexnet}, in its use of stacked 2D
convolutional layers with rectified linear unit (ReLU) activation functions that
alternate with max-pooling layers, and eventually terminated in fully connected
layers with SoftMax outputs.  It differed from AlexNet in terms of
hyperparameters, such as the number of hidden layers, the number of kernels and
their sizes, stride lengths, and dropout values.

The detailed design of the CNN we used is shown in Figure~\ref{fig:cnn}.   The
network is composed of four 2D convolutional layers, each followed by a maximum
pooling layer~\cite{ref:lenet5,ref:alexnet}.   The number of output filters in
the sequence of four convolutional layers are 16, 32, 32, and 64, respectively. 
Filters in all four convolutional layers have kernel sizes of $3\times3$, stride
lengths of one, and use ReLU activation functions.  The pool sizes used in the
pooling layers are $4\times4$ for the first layer and $2\times2$ for all
subsequent layers.  Stride lengths for all pooling layers correspond to their
pool sizes.  The final two layers of the network, following the fourth pooling
layer, are fully connected (FC) layers.  The first FC layer has 128 neurons with
ReLU activation functions and the last FC layer has 2 output neurons using
SoftMax activation functions. The larger of these two outputs, which sum to
a value of one, was selected to determine the model prediction. ``Dropouts" are
performed prior to each FC layer in which a fraction (0.4 and 0.8 for the first
and second FC layers, respectively) of the inputs are randomly ignored.  This
method lessens the chances of overfitting by minimizing co-adaptations between
layers that do not generalize well to unseen data~\cite{ref:dropout}. The total
number of parameters in the model is 1,212,578.

\subsection{Training the Model}
\label{sec:training}

The images used for training the model were derived from $800\times723$ pixel,
8-bit grayscale images in the portable network graphics format, covering the
full DECam focal plane.  These images were produced with the STIFF program
\citep{ref:stiff}, assuming a power-law intensity transfer curve with index
$\gamma = 2.2$. Minimum and maximum intensity values were set to the 0.005 and
0.98 percentiles of the pixel value distribution, respectively. The training set
consisted, initially, of equal portions of images that had ghosts/scattered
light (positives) and images that did not (negatives).   The positive sample
consisted of 2,389 images that the ray-tracing program identified as likely to
have ghosts/scattered light artifacts and was drawn from the full set of
$\sim$132k images from all DES observing periods. After excluding the images
flagged by ray-tracing program, an equal number of images were randomly selected
from the remainder of the full data set to form the negative sample of the
training set.

Prior to feeding the images to the network, they were first downsampled to
$400\times400$ pixels, which is the input size of the first convolutional
layer.  The pixel values in each image were then normalized to a range whose
minimum and maximum corresponded, respectively, to the first quartile $Q_1(x)$
and third quartile $Q_3(x)$ of the full distribution in the image, by
multiplying each pixel value, $x_i$, by a factor 
$s_i=\frac{x_i-Q_1(x)}{Q_3(x)-Q_1(x)}$.  To improve the model's ability to
correctly identify images that contain ghosts/scattered light artifacts, the
training images were also randomly flipped either along the horizontal axis by
reversing the ordering of pixel rows, or along the vertical axes by reversing
the ordering of pixel columns. This was done using the
\texttt{ImageDataGenerator} class in Keras, which does an in-place substitution
of the input images with the flipped versions, without changing the total size
of the data sample~\cite{ref:chollet2015keras}.

\subsubsection{Model Training Procedure}\label{sec:trainingproc}
The model was trained using 80\% of the sample described in the previous section
and the remaining fraction was set aside for validation. Apart from this
training/validation sample was a separate test sample used to evaluate the
model, which is described in Section~\ref{sec:evalmodel}. Optimal weights for
the model were obtained using {\it Adam}~\cite{ref:adam}, a version of the
mini-batch stochastic gradient method that uses dedicated learning rates for
each parameter and adapts their values based on their history.  The weights were
updated iteratively in randomly picked batches of 32 images (\emph{batch size}),
completing a full pass over the entire sample in one \emph{epoch}. A total of 30
training epochs were performed.  The loss function used was categorical
cross-entropy, calculated according to $L=-\sum_{i=1}^N\sum_{j=1}^{M}y_{ij}\cdot
log(p_{ij})$, where the index $i$ runs over the number of observations, $N$, and
the index $j$ is taken over the number of classes, $M$.  $p_{ij}$ is the
probability and $y_{ij}$ is either 0 or 1, depending on whether class $j$ is the
correct classification for observation $i$. In our case, we have two classes
($M=2$) corresponding to whether or not an image contains a ghost/scattered
light artifact.

Upon visual examination of the false positives and false negatives after
training, it was found that some images were mislabeled.  This was because
images labeled as lacking ghosts/scattered light artifacts were initially
selected based on the ray-tracing program output. As it turned out, many
``clean" images actually contained ghosts/scattered light.  When images that
were positively identified by the ray-tracing program were inspected, the
opposite case was also found to be true -- some images labeled as having
ghosts/scattered light did not exhibit detectable artifacts. Therefore, several
iterations were required in order to fix the mislabeled images and repeat the
30-epoch training process. 

\subsubsection{Training and Validation Results}\label{sec:trainvalidresults}

\begin{figure*}[htbp]
\centering
\includegraphics[width=0.5\textwidth]{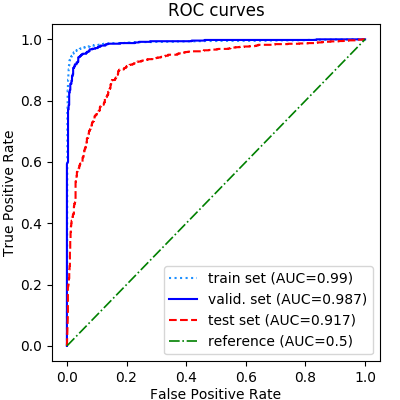}
\caption{ROC curves and the associated areas under the ROC curves (AUCs) are shown separately for the training, validation, and independent test samples. The green dash-dotted line represents the reference case of no discriminating power (AUC=0.5).}
\label{fig:roc}
\end{figure*}

\begin{figure*}[htbp]
\centering
  \begin{tabular}[t]{ccc}
    \includegraphics[width=0.3\textwidth]{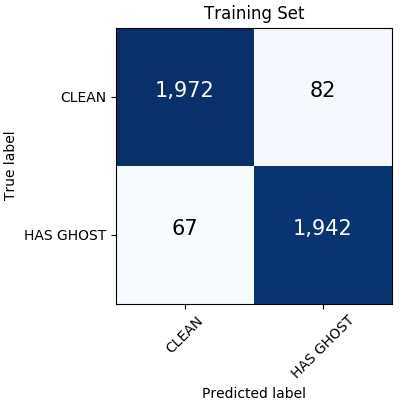} &
    \includegraphics[width=0.3\textwidth]{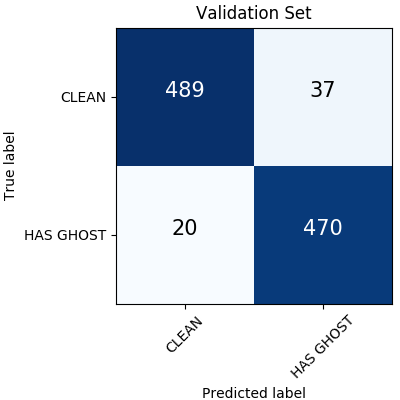} &
    \includegraphics[width=0.3\textwidth]{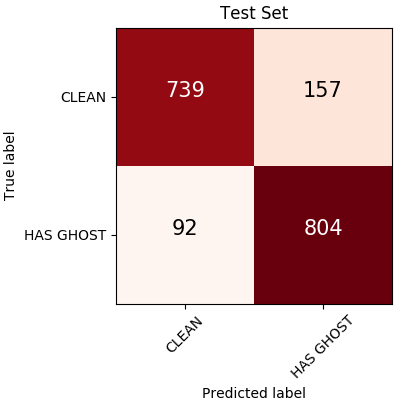} \\
    (a) & (b) & (c) \\
    & & \\
  \end{tabular}
\caption{The confusion matrices are shown separately for the (a) training, (b) validation, and (c) independent test samples. In each matrix, the number of true negatives and positives are shown, respectively, in the upper left and right boxes, while the number of false positives and negatives are shown, respectively, in the upper right and lower left boxes.}
\label{fig:cm}
\end{figure*}

The final results of training are shown in Figures~\ref{fig:accloss},
\ref{fig:roc}, and \ref{fig:cm}. The two panels in Figure~\ref{fig:accloss} show
the evolution of the training accuracy (left) and loss (right) over the epochs. 
The validation curves follow the training curves closely, indicating no
overfitting. Accuracies of over 94\% are achieved on both training and
validation sets at the end of 30 epochs.

\begin{table*}[htb]
    \centering
    {\begin{tabular}{ccccc}
         \hline\noalign{\smallskip}
         \multicolumn{5}{c}{Performance Summary} \\
         Sample & Accuracy & Precision & Recall & AUC \\
         \noalign{\smallskip}\hline\noalign{\smallskip}
         training & 0.963 & 0.959 & 0.967 & 0.990 \\	   	  
         validation & 0.944 & 0.927 & 0.959 & 0.987 \\
         test  & 0.861 & 0.837 & 0.897 & 0.917 \\
         \noalign{\smallskip}\hline
    \end{tabular}}
    \caption{Summary of performance metrics for each sample. Accuracy, precision, and recall are calculated as described in Section~\ref{sec:evalmodel} using the values in Figure~\ref{fig:cm}.  The AUCs are the areas under the ROC curves in Figure~\ref{fig:roc}.}
    \label{tab:metrics}
\end{table*}

Figure~\ref{fig:roc} plots the receiver operating characteristic curve (ROC) for
the trained model, showing the true postive rate versus the false positive rate.
The curves resulting from the application of this model to the training
(light blue dotted line) and validation (solid blue line) samples are shown
separately. The area under the ROC curve (AUC) for the validation sample
is 0.987, indicating good separation between the two classes of images.  For
comparison, the diagonal green dash-dotted line shows the case when a model has
absolutely no discriminating power between classes where AUC=0.5.

Figures~\ref{fig:cm}a~and~\ref{fig:cm}b plot the confusion matrices for the
training and validation samples, respectively. In each matrix, the values in
the first row represent the number of true negatives in the first column and the
number of false positives in the second column. The values in the second row
represent the number of false negatives in the first column and the number of
true positives in the second column.

\subsection{Evaluating the Model}\label{sec:evalmodel}

The validation set was not used directly to train the model, however, it served
as an early indicator of model performance in the training process.  In this
respect, it could have influenced the model and hyperparameter choices.  The
performance of the fully trained model was therefore evaluated in an unbiased
way using an independent test data sample.  This sample was constructed by
visually selecting an equal number of images containing ghosts/scattered light
artifacts and those without them, and labeling them according to their true
class. It consisted of 1,761 DECam images spread across all DES data taking
periods.  It also excluded all the images used for training and validation, and
was $\sim$37\% of that sample in size. The fully trained model was applied to
this sample to predict which class they belonged to.  The ROC curve for the test
data sample is represented by the dashed red line in Figure~\ref{fig:roc} with
AUC=0.917, indicating good discrimination between the two classes.  From the
confusion matrix shown in Figure~\ref{fig:cm}c, one calculates
$accuracy=\mathrm{\frac{TP+TN}{Total}}=0.861$,
$precision=p=\mathrm{\frac{TP}{TP+FP}}=0.837$,
$recall=r=\mathrm{\frac{TP}{TP+FN}}=0.897$, and $F_{1}=2\cdot\frac{p\cdot
r}{p+r}=0.866$, where TP, FP, TN, and FN are, respectively, the number of true
positives, false positives, true negatives and false negatives.  These
results are summarized in Table~\ref{tab:metrics} together with those for the
training and validation samples.

\begin{figure*}[htbp]
\centering
  \begin{tabular}[t]{ccc}
    \includegraphics[width=0.31\textwidth]{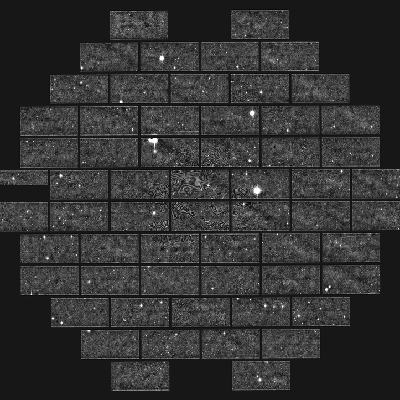} &
    \includegraphics[width=0.31\textwidth]{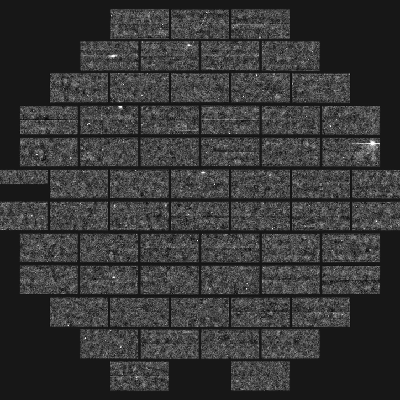} &
    \includegraphics[width=0.31\textwidth]{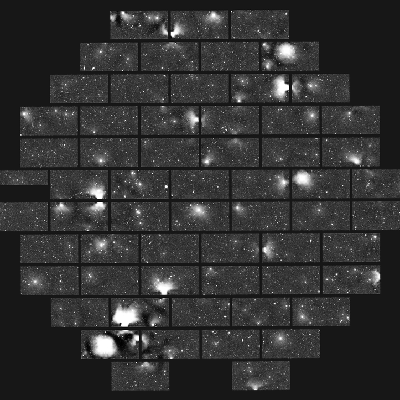} \\
    (a) & (b) & (c) \\
    \includegraphics[width=0.31\textwidth]{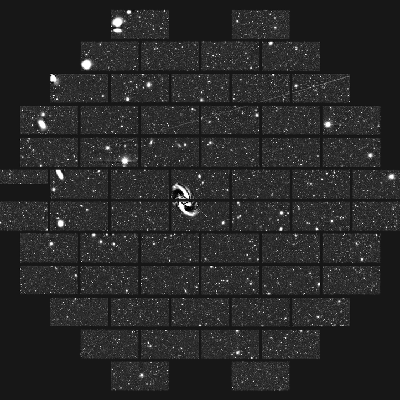} &
    \includegraphics[width=0.31\textwidth]{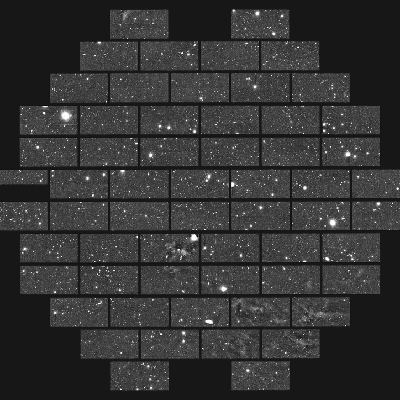} &
    \includegraphics[width=0.31\textwidth]{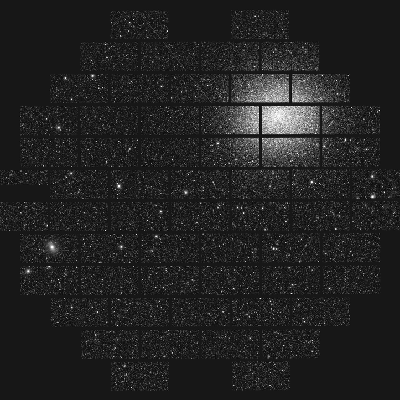} \\
    (d) & (e) & (f) \\
    \includegraphics[width=0.31\textwidth]{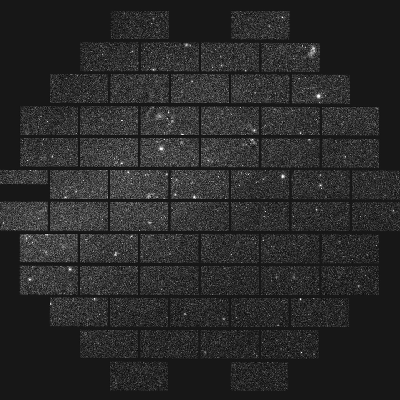} &
    \includegraphics[width=0.31\textwidth]{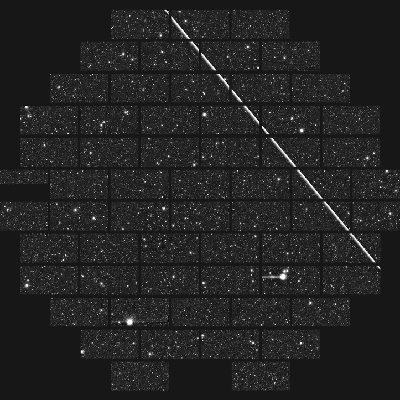} &
    \includegraphics[width=0.31\textwidth]{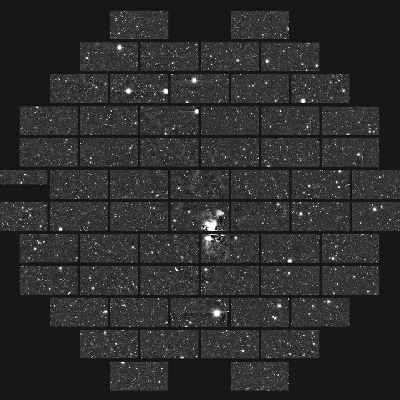} \\
    (g) & (h) & (i)\\
  \end{tabular}
\caption{Example false positives found by the trained model in the test set
described in Section \ref{sec:evalmodel}. The exposures shown in panels (a),
(b), and (c) have poor data quality due to heavy cloud cover  which contributes
to misclassification by the CNN. The barred spiral NGC 1365 in the Fornax galaxy
cluster (d), Galactic cirrus (e), and the Omega Centauri globular cluster in
(f), exhibit features similar to those found in ghosting artifacts. The faint
resolved stars in the periphery of the LMC in (g), and the artificial
earth-orbiting satellite track in (h), have features found in scattered light
artifacts. There is a barely visible ghost artifact in columns 4 \& 5 of the
middle two rows of CCDs in (i).}\label{fig:testfp}
\end{figure*}

\begin{figure*}[htbp]
\centering
  \begin{tabular}[t]{ccc}
    \includegraphics[width=0.31\textwidth]{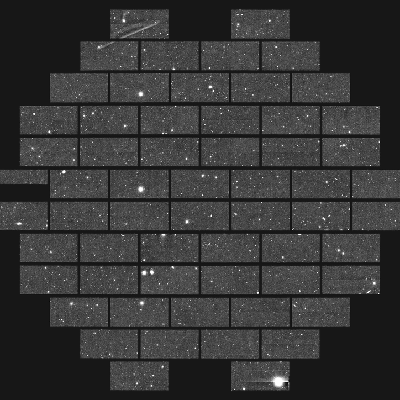} &
    \includegraphics[width=0.31\textwidth]{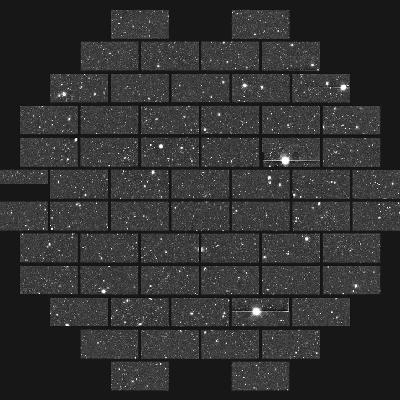} &
    \includegraphics[width=0.31\textwidth]{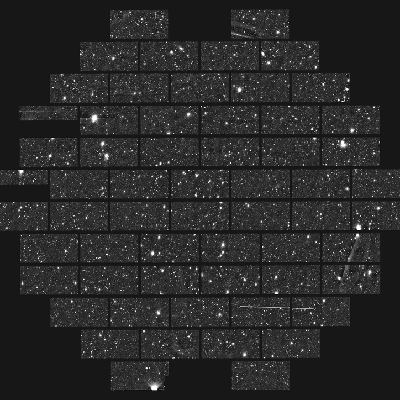} \\
    (a) & (b) & (c) \\
  \end{tabular}
\caption{Selected examples of false negatives found by the trained model in
the test set described in \ref{sec:evalmodel}. Faint ghosts/scattered light
artifacts are visible in the upper left corner of (a), rightmost column CCD in
the 5th row from the top of (b), and rightmost column CCDs in the 8th and 9th
rows from the top of (c). }\label{fig:testfn}
\end{figure*}

Typical examples of misclassified images from the test sample, in the form
of false positives and false negatives, are shown in
Figures~\ref{fig:testfp}~and~\ref{fig:testfn}, respectively.  Although the
images in the first class of false positives represented by
Figures~\ref{fig:testfp}a--\ref{fig:testfp}c do not bear an obvious resemblance
to those containing ghosting/scattered light artifacts, they all exhibit poor
data quality from nearly a magnitude of extinction due to clouds that may be
confusing the CNN.  These images do not pass the high-level DES data quality
criteria.  The second class of false positives contain objects that exhibit
features similar to those found in ghosting artifacts
(Figures~\ref{fig:testfp}d--\ref{fig:testfp}f) and scattered light artifacts
(Figures~\ref{fig:testfp}g \& \ref{fig:testfp}h), making them intuitively easier
to appreciate.  The third class of false positives, represented by
Figure~\ref{fig:testfp}i, are in some sense true positives, because they contain
faint artifacts close to the human detection threshold.  In this image, there is
a ghost artifact faintly visible in the 4th and 5th columns from the left, in
the two middle rows of CCDs.  The false negatives in Figure~\ref{fig:testfn} are
easier to understand because they all contain ghost artifacts that are not too
difficult to see (their locations are described in the figure caption).

Our application involves a large data set where images with ghosts/scattered
light constitute a relatively small fraction of the entire sample.  False
negatives carry a high cost due to their detrimental effects on astronomical
measurement and the difficulty of manual identification in a data set of this
size.  On the other hand, false positives are less of a problem since they are
easier to identify from the smaller sample predicted by the model to be
ghosts/scattered light.  Our model's true positive rate or $recall$ of 
$\sim$90\% shows it is able to identify  a significant fraction of all images
with ghosts/scattered light, and its $precision$ of $\sim$84\% indicates that
false positives are also kept under control, both of which are favorable
characteristics for this application.  As indicated by the AUC, our model
performs better on the training and validation set than on the test set.   This
may be an indication of biases introduced in the construction of the former set,
which is based on images identified by the ray-tracing program.

\begin{figure*}[htbp]
\centering
  \begin{tabular}[t]{ccc}
    \includegraphics[width=0.38\textwidth]{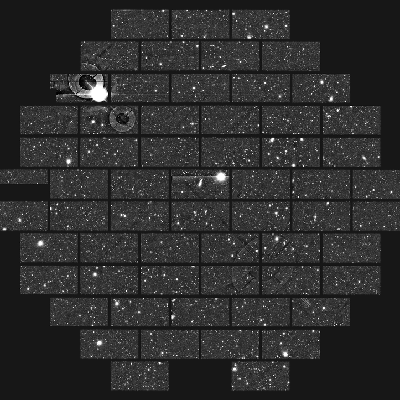} &
    \includegraphics[width=0.38\textwidth]{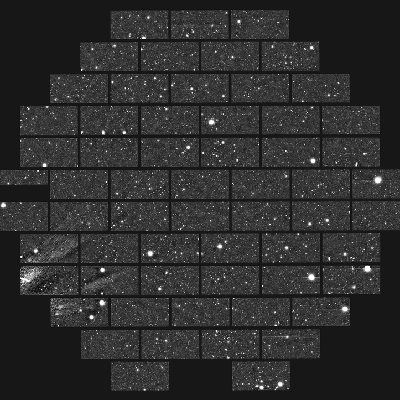} \\
    (a) & (b)\\
    \includegraphics[width=0.38\textwidth]{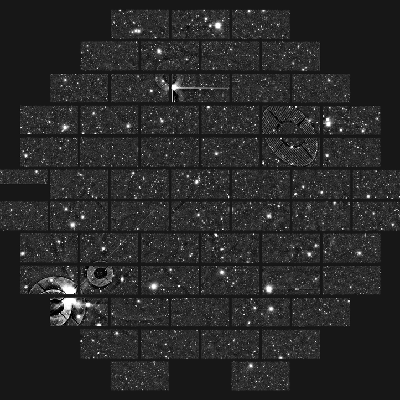} &
    \includegraphics[width=0.38\textwidth]{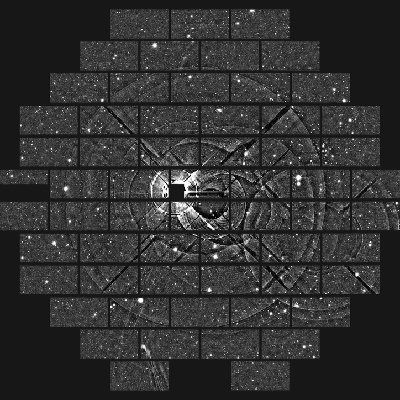} \\
    (c) & (d)\\
    \includegraphics[width=0.38\textwidth]{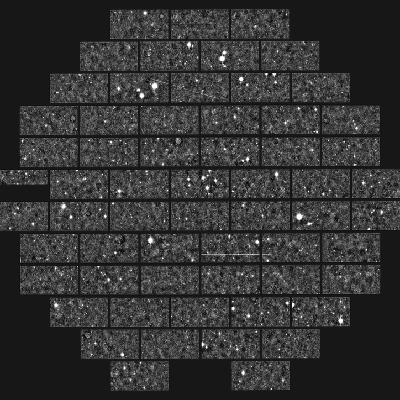} &
    \includegraphics[width=0.38\textwidth]{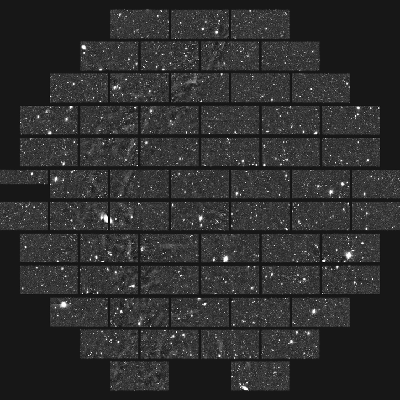} \\
    (e) & (f)\\
  \end{tabular}
\caption{The images above are examples of DES Year 5 images predicted by the CNN
described in this paper to exhibit ghosts/scattered light artifacts, but which
were not identified by the ray-tracing algorithm as such. Figures (a) to (d)
show examples that have actual artifacts, representing true positives.  Figures
(e) and (f) are examples of the $\sim$23\% described in the text that either do
not exhibit artifacts or have negligible levels, representing false
positives.}\label{fig:y5ghosts}
\end{figure*}

\section{Applying the Trained Model on DES Data and Comparing with the Traditional Method}\label{sec:desy5}

The CNN trained according to the details described in Section~\ref{sec:training}
was used to perform inference on the DES Year-5 data set consisting of 23,755
full focal plane DECam images with exposure numbers ranging from 666747 to
724364, which were prepared using the procedure described in
Section~\ref{sec:training}. This set also included the Year-5 images that were
used in the training+validation and testing stages. For each image, the model
was used to predict whether it contained ghosts/scattered light or whether it
was free from such artifacts.  The model identified 3,285 images as
positives, containing ghosts/scattered light artifacts. Several examples of
these images are shown in Figure~\ref{fig:y5ghosts}. Only 716 images in this set
of positives were false positives, exhibiting nearly imperceptible or no sign of
ghosts/scattered light artifacts.  The precision achieved was therefore
$p=\nicefrac{2569}{3285}=0.782$.

For comparison, the ray-tracing program described in
Section~\ref{sec:conventional} classified 259 DES Year-5 images as
containing artifacts. Out of these, 241 were in common with the set of positives
identified by the ML model, and all of the images in this overlap region were
true positives.  The remaining 18 that were positively classified only by the
ray-tracing program were all true positives except for 8. The precision achieved
by the ray tracing model was therefore $p=\frac{{241+10}}{259}=0.969$.

The difference in precision from the two methods may be due to the more limited
range of image types dealt with by the ray-tracing program, and the issue raised
in Section~\ref{sec:evalmodel} about the training and validation set being based
on the images identified by that program.

\section{Computer Resource Utilization}

The conventional ray tracing algorithm takes on the order of a few ms per
image for actual ray tracing. Additional time is spent querying the bright star
catalog around each exposure as a pre-processing step. This algorithm was run on
a yearly basis as input to the DES data processing.

For the CNN-based approach, training the model over 30 epochs using the
procedure described in Section~\ref{sec:trainingproc} on a laptop with an Intel
Xeon E-2176M CPU, 32GB RAM, and a mid-range 4GB Nvidia Quadro P2000 Mobile GPU
took 8.8 min (18 s/epoch) to complete. Utilizing the 16GB Nvidia P100 GPUs
available in the Google Cloud Colaboratory Jupyter notebook
environment~\cite{ref:colab}, reduces the training time by a factor of $4\times$
(4.4 s/epoch).

The process of performing inference with the CNN on the 23,755 image DES
Year-5 data set described in Section~\ref{sec:desy5} took 50 s (2 ms/image) on
the Quadro-equipped laptop described above.  Such short inference times are
indeed promising for real-time artifact identification on future large-scale
cosmic surveys, especially since the network model has not even been optimized
for speed yet.  Furthermore, there now exist practical high-level synthesis
tools that can implement these network models on FPGA hardware for critical
real-time applications~\cite{ref:Duarte_2018ite}.

\section{Conclusion}

We have successfully applied a machine learning based method to identify DES
images containing ghosts/scattered light artifacts.  This method
positively identified $\sim$97\% of all images that had been
previously identified as containing artifacts by a traditional ray-tracing
method.   Overall, it also identified $\sim$10$\times$ more images with actual
artifacts, with a precision of $\sim$78\%. This serves as a proof-of-principle
demonstrating the effectiveness of using modern ML methods in identifying
ghosts/scattered light in optical telescope images from a cosmic survey.  It
lays the foundation for possible future refinements. The scope of this work
was limited to detecting the presence of these artifacts in an image without
identifying their location within the image. In future work, we will take
advantage of recent developments in object detection and semantic segmentation
to expand the capability of our method to include the identification of the
individual pixels associated with each artifact~\cite{ref:he2018mask}. Such
enhancements, coupled with the results presented in this work, will benefit
future cosmic surveys like the LSST, which will be faced with the challenge of
even larger data sets.

\section{Acknowledgements}

This collaborative work was carried out as part of an Illinois Mathematics and
Science Academy (IMSA) Student Inquiry and Research (SIR) project.  We wish to
thank Dr. Don Dosch, Dr. David Devol, and Dr. Eric Smith of IMSA for overseeing
the SIR program and making this collaboration possible.  We also wish to thank
the staffs of Fermilab's experimental astrophysics group and IMSA's SIR office
for their support.

Funding for the DES Projects has been provided by the U.S. Department of Energy,
the U.S. National Science Foundation, the Ministry of Science and Education of
Spain, the Science and Technology Facilities Council of the United Kingdom, the
Higher Education Funding Council for England, the National Center for
Supercomputing Applications at the University of Illinois at Urbana-Champaign,
the Kavli Institute of Cosmological Physics at the University of Chicago, the
Center for Cosmology and Astro-Particle Physics at the Ohio State University,
the Mitchell Institute for Fundamental Physics and Astronomy at Texas A\&M
University, Financiadora de Estudos e Projetos, Funda{\c c}{\~a}o Carlos Chagas
Filho de Amparo {\`a} Pesquisa do Estado do Rio de Janeiro, Conselho Nacional de
Desenvolvimento Cient{\'i}fico e Tecnol{\'o}gico and the Minist{\'e}rio da
Ci{\^e}ncia, Tecnologia e Inova{\c c}{\~a}o, the Deutsche Forschungsgemeinschaft
and the Collaborating Institutions in the Dark Energy Survey. 

The Collaborating Institutions are Argonne National Laboratory, the University
of California at Santa Cruz, the University of Cambridge, Centro de
Investigaciones Energ{\'e}ticas, Medioambientales y Tecnol{\'o}gicas-Madrid, the
University of Chicago, University College London, the DES-Brazil Consortium, the
University of Edinburgh, the Eidgen{\"o}ssische Technische Hochschule (ETH)
Z{\"u}rich, Fermi National Accelerator Laboratory, the University of Illinois at
Urbana-Champaign, the Institut de Ci{\`e}ncies de l'Espai (IEEC/CSIC), the
Institut de F{\'i}sica d'Altes Energies, Lawrence Berkeley National Laboratory,
the Ludwig-Maximilians Universit{\"a}t M{\"u}nchen and the associated Excellence
Cluster Universe, the University of Michigan, the National Optical Astronomy
Observatory, the University of Nottingham, The Ohio State University, the
University of Pennsylvania, the University of Portsmouth, SLAC National
Accelerator Laboratory, Stanford University, the University of Sussex, Texas
A\&M University, and the OzDES Membership Consortium.

Based in part on observations at Cerro Tololo Inter-American Observatory,
National Optical Astronomy Observatory, which is operated by the Association of
Universities for Research in Astronomy (AURA) under a cooperative agreement with
the National Science Foundation.

The DES data management system is supported by the National Science Foundation
under Grant Numbers AST-1138766 and AST-1536171. The DES participants from
Spanish institutions are partially supported by MINECO under grants
AYA2015-71825, ESP2015-66861, FPA2015-68048, SEV-2016-0588, SEV-2016-0597, and
MDM-2015-0509, some of which include ERDF funds from the European Union. IFAE is
partially funded by the CERCA program of the Generalitat de Catalunya. Research
leading to these results has received funding from the European Research Council
under the European Union's Seventh Framework Program (FP7/2007-2013) including
ERC grant agreements 240672, 291329, and 306478. We  acknowledge support from
the Brazilian Instituto Nacional de Ci\^encia e Tecnologia (INCT) e-Universe
(CNPq grant 465376/2014-2).

This manuscript has been authored by Fermi Research Alliance, LLC under Contract
No. DE-AC02-07CH11359 with the U.S. Department of Energy, Office of Science,
Office of High Energy Physics.

\section{Contributions}
The author list is ordered alphabetically. C.~Chang was responsible for the
initial idea of applying ML techniques to artifact detection and served as
primary mentor to D.M.~Wang in the SIR project.   A.~Drlica-Wagner identified
optical ghosts/scattered light detection as an important problem that could
benefit from an alternative approach.  He also ran the ray-tracing algorithm,
provided access to labeled images containing ghosts/scattered light, and advised
on their classification and causes.  S.~Kent developed the DECam ray-tracing
algorithm and provided expertise on the origins of ghosts and scattered light in
the DECam system.  B.~Nord provided initial training on ML algorithms and
advised on algorithm design.  He also served as the primary liason between the
astrophysics group at Fermilab and IMSA.   D.M.~Wang was responsible for
extending the sample ML models provided by B.~Nord to develop the models used in
this work.  She also performed the training and validation of the model,
including selecting its set of hyperparameters.  M.H.L.S.~Wang prepared the
samples used for training and comparisons, based on the data A.~Drlica-Wagner
provided access to.  He also advised on the preprocessing of the data and
prepared the initial version of this document, including the model architecture
diagram.

\bibliography{desghostbuster_refs}
\end{document}